\documentclass[sigplan,10pt,authorversion]{acmart}

\usepackage{tikz}
\usepackage{xcolor}
\usepackage{xspace}
\usetikzlibrary{shapes.geometric,shapes.multipart,shapes,fit,calc,positioning,patterns,arrows.meta,backgrounds,decorations.pathmorphing,decorations.shapes,decorations.pathreplacing}

\usepackage[noabbrev, capitalise]{cleveref}
\usepackage{listings}

\usepackage{siunitx}

\definecolor{nDarkGrey}{RGB}{152,164,174}
\definecolor{nGrey}{RGB}{210,213,215}
\definecolor{nLightGrey}{RGB}{235,236,238}

\definecolor{nDarkRed}{RGB}{141,20,41}
\definecolor{nRed}{RGB}{201,169,147}
\definecolor{nLightRed}{RGB}{237,231,222}

\definecolor{nDarkGreen}{RGB}{0,155,119}
\definecolor{nGreen}{RGB}{170,207,189}
\definecolor{nLightGreen}{RGB}{229,239,234}

\definecolor{nDarkBlue}{RGB}{0,56,101}
\definecolor{nBlue}{RGB}{144,167,198}
\definecolor{nLightBlue}{RGB}{221,229,240}

\definecolor{nDarkCyan}{RGB}{0,177,235}
\definecolor{nCyan}{RGB}{180,214,245}
\definecolor{nLightCyan}{RGB}{234,243,252}

\definecolor{nDarkYellow}{RGB}{201,147,19}
\definecolor{nYellow}{RGB}{217,198,137}
\definecolor{nLightYellow}{RGB}{243,238,223}

\definecolor{Blue}{RGB}{0,119,187} % vibrant
\definecolor{Green}{RGB}{34,136,51}
\definecolor{DarkGreen}{RGB}{34,85,34}
\definecolor{Cyan}{RGB}{51,187,238}
\definecolor{Teal}{RGB}{0,153,136}
\definecolor{Yellow}{RGB}{221,170,51} % high contrast
\definecolor{Orange}{RGB}{238,119,51}
\definecolor{Red}{RGB}{204,51,17}
\definecolor{DarkRed}{RGB}{102,51,51}
\definecolor{Magenta}{RGB}{238,51,119}
\definecolor{Purple}{RGB}{170,51,119}
\definecolor{Gray}{RGB}{187,187,187}
\definecolor{DarkGray}{RGB}{85,85,85}
\definecolor{Black}{RGB}{0,0,0}
\definecolor{White}{RGB}{255,255,255}
\definecolor{User}{RGB}{34,136,51}
\definecolor{eBPF}{RGB}{170,51,119}
\definecolor{Kernel}{RGB}{0,119,187}
\definecolor{Hardware}{RGB}{102,51,51}

\definecolor{meeting210730}{RGB}{0,0,0}
\definecolor{stefanshighlights210727}{RGB}{0,0,0}
\definecolor{plos21reviews}{RGB}{0,0,0}

\def\xunit{0.67cm}
\def\yunit{0.62cm}
\tikzset{font=\sffamily}

\usepackage[]{acro}
\DeclareAcronym{VM}{short = VM, long = virtual machine}
\DeclareAcronym{JIT}{short = JIT, long = just-in-time}
\DeclareAcronym{IO}{short = IO, long = input/output}
\DeclareAcronym{AIO}{short = AIO, long = asynchronous input/output}
\DeclareAcronym{BPF}{short = BPF, long = Berkeley Packet Filter}
\DeclareAcronym{eBPF}{short = eBPF, long = extended Berkeley Packet Filter}
\DeclareAcronym{KPTI}{short = KPTI, long = Kernel Page Table Isolation}
\DeclareAcronym{iTLB}{short = iTLB, long = instruction address translation lookaside buffer}
\DeclareAcronym{DVFS}{short = DVFS, long = dynamic voltage and frequency scaling}
\DeclareAcronym{vDSO}{short = vDSO, long = a virtual Dynamically-linked Shared Object}
\DeclareAcronym{OS}{short = OS, short-plural = es, long = operating system}
\DeclareAcronym{KASLR}{short = KASLR, long = kernel address space layout randomization}

\newcommand{\eg}{e.g.,~}

\newcommand*{\system}{\textsc{AnyCall}\xspace}
\newcommand*{\manycall}{\system}
\newcommand*{\manycalls}{\textsc{AnyCalls}\xspace}

\newcommand{\userKernelTransition}{user\slash{}kernel transition} % app in user mode <=> os in kernel mode
\newcommand{\titleUserKernelTransitions}{User/Kernel Transitions}
\newcommand{\userKernelTransitions}{user\slash{}kernel transitions}
\newcommand{\transitionOverheads}{transition overheads} % between user/kernel or eBPF/kernel
\newcommand{\bytecodeExecutor}{bytecode executor} % aka 'bytecode VM', 'runtime environment'
\newcommand{\systemCallX}{system-call}  % e.g. 'system-call overhead/rate'
\newcommand{\manyCallFindMagicSpeedup}{\SI{24}{\percent}} % 1-(484ms/633ms)=0.235% Attention: not used in abstract

\acmYear{2021}\copyrightyear{2021}
\setcopyright{acmlicensed}
\acmConference[PLOS '21]{11th Workshop on Programming Languages and Operating Systems}{October 25, 2021}{Virtual Event, Germany}
\acmBooktitle{11th Workshop on Programming Languages and Operating Systems (PLOS '21), October 25, 2021, Virtual Event, Germany}
\acmPrice{15.00}
\acmDOI{10.1145/3477113.3487267}
\acmISBN{978-1-4503-8707-1/21/10}

\begin{document}

%\title{System: Efficient In-Kernel System-Call Chaining}
%\title{System: Rapid System Calls with an In-Kernel Virtual Machine}
%\title{System: Rapid System-Call Chains with an In-Kernel Virtual Machine}
%\title{System: Reducing System-Call Latency with an In-Kernel Virtual Machine}
%\title{System: Reducing System-Call Overhead with an In-Kernel Virtual Machine}
%\title{System: Reducing Operating-System Overhead by In-Kernel System-Call Chaining} %ROSS; SIS, SIC
%\title{\textsc{EOS}: Efficient Operating System Interfaces by In-Kernel System-Call Chaining} %SIS, SIC
%\title[Very Fast System Calls]{Reducing System-Call-Induced Processor Mode Switches
%  with an In-Kernel JIT Bytecode Compiler}
%\title[sys¿all]{sys¿all: Safe, Fast, and Flexible System-Call Aggregation}
%\title[AnyCall]{ManyCall: Safe, Fast, and Flexible System-Call Aggregation}
%\title[AnyCall]{ManyCall: Safe, Fast, and Flexible System-Call Aggregation}
% * system-call clusters/clustering
% \title[AnyCall]{Fast, \textit{not} Furious: System-Call Aggregation with AnyCall}
\title[AnyCall: Fast and Flexible System-Call Aggregation]{AnyCall: Fast and Flexible System-Call Aggregation}

% alt: Very Fast System Calls using \ac{eBPF}
%\title[eBPF System Calls]{Systems Calls {\color{red}without} Processor Mode Switches using
%  the eBPF In-Kernel Virtual Machine}

\def\fau{%
 \institution{Friedrich-Alexander-Universit{\"a}t Erlangen-N{\"u}rnberg}%
 \country{Germany}%
}
\def\rub{%
 \institution{Ruhr University Bochum}%
 \country{Germany}%
}
\author{Luis Gerhorst}
\affiliation{\fau}
\author{Benedict Herzog}
\affiliation{\rub}
\author{Stefan Reif}
\affiliation{\fau}
\author{Wolfgang Schr\"oder-Preikschat}
\affiliation{\fau}
\author{Timo H\"onig}
\affiliation{\rub}

\begin{abstract}
Operating systems rely on system calls to allow the controlled communication of isolated processes with the kernel and other processes.
% 'mode switch' ok here because we want to explain why it's expensive at hardware level
Every system call includes a processor mode switch from the unprivileged user mode to the privileged kernel mode.
Although processor mode switches are the essential isolation mechanism to guarantee the system's integrity, they induce direct and indirect performance costs as they invalidate parts of the processor state.
In recent years, high-performance networks and storage hardware has made the
\userKernelTransition{} overhead the bottleneck for IO-heavy applications.
To make matters worse, security vulnerabilities in modern processors (\eg
Meltdown) have prompted kernel mitigations that further increase the transition overhead.
To decouple system calls from \userKernelTransitions{} we propose \system,
which uses an in-kernel compiler to execute safety-checked user bytecode in kernel
mode.
This allows for very fast system calls interleaved with error checking and
processing logic using only a single \userKernelTransition{}.
We have implemented \system based on the Linux kernel's \ac{eBPF} subsystem.
Our evaluation demonstrates that system call bursts are up to 55 times faster
using \system and that real-world applications can be sped up by
\manyCallFindMagicSpeedup{} even if only a minimal part of their code is run by
\system.
\end{abstract}

%\begin{CCSXML}
%<ccs2012>
%  <concept_id>10010520.10010553.10010562</concept_id>
%  <concept_desc>Computer systems organization~Embedded systems</concept_desc>
%  <concept_significance>500</concept_significance>
%  <concept_id>10010520.10010575.10010755</concept_id>
%  <concept_desc>Computer systems organization~Redundancy</concept_desc>
%  <concept_significance>300</concept_significance>
%  <concept_id>10010520.10010553.10010554</concept_id>
%  <concept_desc>Computer systems organization~Robotics</concept_desc>
%  <concept_significance>100</concept_significance>
%  <concept_id>10003033.10003083.10003095</concept_id>
%  <concept_desc>Networks~Network reliability</concept_desc>
%  <concept_significance>100</concept_significance>
%</ccs2012>
%\end{CCSXML}
%
%\ccsdesc[500]{Computer systems organization~Embedded systems}

% \keywords{eBPF, Linux, Operating Systems/Kernels, Performance}

\settopmatter{printacmref=false}

\maketitle

\acresetall{}

\section{Introduction}
\label{sec:introduction}

% {\color{gray}``Six or fewer pages of technical content, including all text, figures, tables, appendices, etc., excluding references.''}

% {\color{gray}``Reviewing will be double-blind: no author or affiliation information should appear in the submission.''}

General-purpose computing systems rely on memory isolation as the most fundamental security and safety mechanism that enables protection from malicious activities, maintains privacy, and confines faulty programs~\cite{aiken:2006:mspc,ge:2019:eurosys}.
The necessary counterpart to memory isolation is a well-defined communication interface that provides kernel-level functionality safely to user programs, typically implemented by \emph{system calls}.
The costs of system calls, in particular their execution-time overhead, have been a well-known performance-critical system property for decades~\cite{elphinstone:2013:sosp}.
Besides the traditional \emph{direct} costs of isolation, each transition
between user and kernel space inflicts \emph{indirect} costs as the processor
state must be (partially) invalidated~\cite{soares:2010:osdi}.
% Evaluations on easy16 has shown more branch / l1_icache misses for the
% ManyCall variant, with/without mitigations:
%
% {\color{red}Our experiments indicate that this includes parts of the branch target buffer and L1
% cache on recent AMD Zen 2 processors, causing the number of cache misses to
% scale with the system call rate.}
%
The importance of such indirect costs is expected to grow since caches and hardware buffers are becoming increasingly performance-critical.
With the discovery of Meltdown~\cite{lipp:2018:meltdown} and Spectre~\cite{kocher:2019:spectre}
%, and further processor-level timing side-channels,
the costs of system calls have increased even further, often causing a significant overhead on execution time~\cite{prout:2018:measuring,ren:2019:sosp} and energy demand~\cite{herzog:2021:eurosec}.
The required flushes of processor-internal buffers enforce isolation, but cause significant overheads, particularly for applications that execute system calls at a high rate.

An example for \systemCallX{}-intensive applications is the Unix \texttt{find}
tool that traverses directories and filters files by configurable criteria.
{\color{plos21reviews} Therefore, numerous system calls are executed to read the
  file-system tree, but the amount of computation in user space, in-between
  system calls, is often negligible.}
% (e.g., checking the file's owner).
Experiments on a \SI{1.8}{\GHz} {\color{meeting210730}desktop computer} show
that
% tar xf: 430620/6.182s=69_657 syscalls/s
% find: 64325/0.092s=699_184 syscalls/s
% du -s: 109698/0.151s=726_476 syscalls/s
\num{70 000} to \num{726 000} system calls are performed per second when
unpacking archives, listing files, and estimating disk usage using the GNU
coreutils\footnote{Numbers obtained by running Debian 10's \texttt{tar xf},
  \texttt{find}, and \texttt{du -s} on the Linux 5.0 source tree. See
  \cref{sec:evaluation} for details on the evaluation setup.}.

Considering that applications with high \systemCallX{} rate suffer most from
Meltdown-type mitigations, we propose \system, a system where only a single
\userKernelTransition{} is required for an arbitrary number of system calls. To
reduce the number
of transitions between user and kernel space, the control flow migrates to kernel
space, calls arbitrary system calls interleaved with application-specific logic,
and eventually returns to user space. Isolation of the supplied application logic
is enforced by an in-kernel \bytecodeExecutor{} (i.e., eBPF) and static bytecode
analysis.

The contributions of this paper are three-fold:
\begin{itemize}
  \item We present an approach to decouple the number of \userKernelTransitions{}
        from the number of system calls, while maintaining isolation using an in-kernel
        \bytecodeExecutor{}.
  \item We extend Linux to support system calls from within \acs{eBPF} programs.
        For this, \system enables verifiably safe access to user memory inside
        eBPF programs to construct system call arguments and access results.
  \item We evaluate \system by comparing the overheads of eBPF-based to traditional system calls.
\end{itemize}

{\color{plos21reviews} This paper is structured as follows.}
\cref{sec:background} presents related work and background knowledge.
\cref{sec:design} discusses the design of \system{} followed by
\cref{sec:implementation} presenting our efficient \acs{eBPF}-based
implementation for the Linux kernel. In \cref{sec:evaluation} we evaluate
\system{} using real-world applications and microbenchmarks. Future work is
discussed in \cref{sec:futurework}. Finally, \cref{sec:conclusion} concludes
this paper.

\section{Background and Related Work}
\label{sec:background}

This section discusses recent hardware and software developments related to the overheads of system calls. We also review existing
techniques that avoid the overhead of \userKernelTransitions{} and give a
brief introduction to \ac{eBPF}.

\subsection{High-Performance \acs{IO}}\label{sec:io}
% Message: Mode switching overhead becomes more relevant because of high
% performance IO.

% TODO: compile into a paragraph

% \begin{itemize}
%   \item Kernel-bypass networking~\cite{belay:2014:osdi,peter:2014:osdi}
%   \item HPC
%   \item System call clustering
% \end{itemize}

% Disks get faster.
%
% APSys'21 Review #16A aggrees: This paper is targeting an important problem
% because the overhead of system calls becomes relatively larger as a result of
% fast devices like NVM.
Traditional hard drives with high access latencies are now superseded by
low-latency media such as NVME-SSDs and Optane storage. This device latency
reduction makes the \userKernelTransition{} delay more significant for the
overall performance of a computing
system~\cite{enberg:2019:hotos,zhong:2021:bpf}. Similarly, networking hardware
is bottlenecked by the software
stack~\cite{rumble:2011:hotos,kaufmann:2019:eurosys}. Therefore, network-heavy
applications directly access network hardware, bypassing the OS
Kernel~\cite{belay:2014:osdi,peter:2014:osdi}.

% Zhong et al. display the latency per storage access layer in table 1. The
% kernel crossing only makes up 5.6 percent but this does not account for the
% indirect cost which affects every software layer below.

% Many Syscalls in Unix File Accesses, already covered by Stefan's example at
% the end of the intro:
%
% When files are cached in kernel memory, the mode switches required to access
% them are a major contributor to the run time of the application. Evaluation of
% common Unix applications (e.g.,~\texttt{find}) shows that they perform up to
% one million mode switches per second while only spending a short amount of
% time in user space continuously (usually less than 1us). This shows that a
% mechanism that can be used to avoid these mode switches would have a large
% impact on the performance of the application. Also, the small amount of time
% spent in user space suggests that the code executed there is not very complex
% (e.g., it only performs error checks and then requests the system call that is
% to be executed next).

\subsection{Meltdown and Spectre}\label{sec:meltdown}

% APSys'21 Review #16B agrees: It is still important to reduce the overhead of
% system-call invocation because of the mitigation against new attacks such as
% Meltdown.
Recently, the Meltdown~\cite{lipp:2018:meltdown} and
Spectre~\cite{kocher:2019:spectre} hardware vulnerabilities have made it
necessary to develop hard\-ware, firm\-ware, and software mitigations. As these
vulnerabilities can circumvent the memory isolation between processes, software
mitigations at \ac{OS} level have been developed. Although these mitigations are
effective in fully or at least partly preventing the exploitation of these
vulnerabilities, they come with potentially significant execution time and
energy demand
overheads~\cite{herzog:2021:eurosec,alhubaiti:2019:impact,prout:2018:measuring}.

% \begin{itemize}
%   \item KPTI also offers security benefits if the processor is not affected by
%         meltdown: KASLR can be bypassed on some AMD processors if KPTI is not
%         active \cite{hund:2013:kaslr}
%   \item \url{https://docs.clip-os.org/clipos/kernel.html#configuration}
%         CLIP-OS argues you
%         should use KPTI to generally reduce hardware side channels.
%   \item KPTI also prevents the ret2user control flow hijacking attack if
%         hardware does not prevent it
%         https://github.com/a13xp0p0v/linux-kernel-defence-map/blob/master/linux-kernel-defence-map.svg
% \end{itemize}
Especially Linux' mitigation against Meltdown and {\color{meeting210730} attacks
  bypassing \acs{KASLR}~\cite{hund:2013:kaslr}}, that is \emph{Kernel Page Table
  Isolation}~(KPTI), can introduce significant overheads for
\userKernelTransitions{}. The most important reason for this are additional TLB
flushes before switching to user space.
%, which can be reduced by the use of the \emph{Process Context Identifiers} (PCID), but not fully avoided.
%Furthermore, PCID requires specific CPU features and at least Linux kernel version v4.14.
The mitigations' overhead for different variants of Spectre attacks, apply more
selectively depending on the workload, but can nevertheless be
significant~\cite{herzog:2021:eurosec}.

\subsection{Avoiding \titleUserKernelTransitions{}}\label{sec:avoid}

In special cases, \ac{AIO} APIs, multicall interfaces, or \ac{vDSO}
can be used to avoid \userKernelTransitions{}.

% Async. IO is limited.
With asynchronous APIs, user space can submit many jobs to the kernel using a single
\userKernelTransition{}~\cite{bhattacharya:2003:aio,soares:2010:osdi,axboe:2019:iouring,corbet:2019:iouring}.
%Furthermore, \ac{AIO} is limited as dependencies between
%operations still require a switch to user mode to execute application logic.
%This can be prevented by using multiple cores (one running the application in
%user mode and one the \ac{OS} in kernel mode), but then the jobs have to be
%communicated using expensive inter-core communication.
%For example, applications can use \texttt{io\_uring} to issue system calls by writing a request to an in-memory buffer that is read by a concurrent thread on another core, running in Kernel mode.
Asynchronous interfaces reduce the \systemCallX{} overhead as the transition
latency can be hidden if an in-kernel thread runs in parallel to the application
(i.e., exploiting parallelism). However, the application-specific logic that
issues system calls and processes their results runs in user space, requiring
data to be communicated between kernel and user space frequently. Multicall
interfaces, as implemented by the Xen hypervisor, also suffer from this
limitation, constraining their practical use to batched page-table updates and
networking hardware control~\cite{pan:2011:hypervisor}. Furthermore, using
\ac{AIO} efficiently requires a programming model many developers are not
familiar with~\cite{atlidakis:2016:eurosys}.
% Also: AIO/io_uring require modifying the system call implementation in order to be
% efficient.

In Linux, some system calls only read a small amount of information. They can be
implemented using \ac{vDSO}~\cite{linux:2021:vdso}, where the data is mapped
into user space, making it directly readable without a processor mode
switch. % 'mode switch' ok here because we just said user space
However, \ac{vDSO} is limited to read-only data for security reasons.

\subsection{\Acl{eBPF}~(\acs{eBPF})}\label{sec:ebpf}

% classical BPF
Initially, the \ac{BPF} was developed to filter network packets directly in the
Linux kernel without having to pass the data to user
space~\cite{mccanne:1993:bpf}. \textcolor{plos21reviews}{This could also be
  achieved by loading custom modules into the kernel, but using \ac{eBPF} provides isolation and is more portable.}
The original \ac{BPF} instruction set is very limited
(\eg no loops) and intentionally not Turing complete.
% eBPF, Performance
Extended \ac{BPF} (\acs{eBPF}) is also not Turing complete (i.e., guaranteed termination), but has a
redesigned instruction set optimized for C interoperability and
compilation to native machine code. Also, recent versions allow for safely bounded
loops. On most architectures, \ac{eBPF} instructions directly map to
machine instructions.
%
% --> end of section
%\ac{eBPF} is currently being ported to other \acp{OS} besides
%Linux~\cite{microsoft:2021:ebpf}.

% Applications
\ac{eBPF} allows applications to inject small code fragments as event handlers in the Linux kernel.
It thus enables detailed but flexible configurability with relatively little overhead.
The applications of \ac{eBPF} include
packet filtering~\cite{miano:2018:network}, tracing~\cite{iovisor:2021:bcc},
\texttt{seccomp} access-control policies~\cite{drewry:2012:seccomp}, file system
sandboxing~\cite{bijlani:2018:sandfs}, caching~\cite{ghigoff:2021:bmc}, and
paravirtualization~\cite{amit:2018:hyperupcalls}.

% Security
% already mentioned
%Starting 2019, bounded loops are also permitted.
%\ac{eBPF} programs can call helper functions
%offered by the kernel to perform complex tasks that would not pass the static
%analysis. These also enable communication with user space and with other
%\ac{eBPF} programs over special data structures called \emph{maps}. Maps are
%concurrent key/value dictionaries synchronized and allocated by the kernel.

% load, attach, types
Registration of an \ac{eBPF} program involves multiple steps. First, the kernel loads the \ac{eBPF} program and analyzes it to guarantee memory safety and a bounded execution time.
Second, the \ac{eBPF} program is compiled from bytecode to native machine code and attached to an event (e.g., tracepoints).
In the following, the already-compiled \ac{eBPF} program is invoked whenever the corresponding event triggers.

As of July 2021, the use of \ac{eBPF} does not support the execution of code from
unprivileged sources for security reasons~\cite{lwn:2019:unprivileged}.
% TODO: we expect that meltdown/spectre can be exploited from within a eBPF program
However, there exists a kernel capability which grants processes not running as
\texttt{root} the ability to load and execute \ac{eBPF} programs.
% TODO: trusted systems services can use it
% TODO: Details
%Recently, Microsoft has published an open source \ac{eBPF} implementation that runs on top
%of Windows~\cite{microsoft:2021:ebpf}.
%Our approach is not limited to Linux, hence \system{} can be easily ported to Windows utilising its \ac{eBPF} implementation.

% The maintainer of the \texttt{io\_uring} asynchronous I/O API is working on
% integrating \ac{eBPF} into the API. This will give users an asynchronous,
% event-based interface to the system calls supported by
% \texttt{io\_uring}~\cite{axboe:2020:plan}.

\section{Design}
\label{sec:design}

%\cref{fig:bpftasks} compares the to system calls from user space. Executing
%three consecutive systems calls from user space incurs a high direct and indirect
%cost. Executing the same three system calls using \ac{eBPF} incurs only one
%third of the direct mode switching overhead and speeds up both user and kernel
%code by not flushing processor state on each system call.

This section presents the design of \system which allows for fast and flexible
system call aggregation using an in-kernel \bytecodeExecutor{}. We discuss the
usage, execution model, and present methods for making \manycall available to
unprivileged processes.

\subsection{Usage}

\textcolor{plos21reviews}{
% Old: The effort required to adopt \system is small, even when using only our prototype tooling.
To keep the adoption effort small, the programming model of \system
must be straight-forward even when using our prototype tooling.
%
% C code with high system call rates can be moved to \manycall{} by
To move userspace C code with high system call rates to \system, a programmer
% (1) replacing it with an \manycall{} and
(1) embeds the code into our framework, compiling it to \ac{eBPF} bytecode, and
(2) replaces the original code with a single system call to invoke the newly
created \manycall.
The \ac{eBPF} runtime of the Linux kernel allows for very expressive aggregation
programs and is extended continuously while our \texttt{libbpf}-based framework
provides transparent \texttt{libc} system call declarations. Therefore,
modifying the code is only required rarely, for example, when system-call
results are to be accessed by reference or when loops are unbounded.
An example \ac{eBPF} program with these issues resolved is displayed in
\cref{fig:effort}. Besides this, one only has to be wary of program size to not
exceed the analyzer's limits on \ac{eBPF} instructions and control flow
complexity. However, these can be easily circumvented by splitting code into
multiple \manycalls.}
%
% But even then, porting is {\color{red}straightforward: Insert helper
%   calls to copy the system-call results from user memory and transform the
%   complex loops into bounded loops that are invoked repeatedly} from outside the
% \manycall{}.
%

\textcolor{plos21reviews}{
Both detection and transformation of code suited for \system{} can be performed by modern compilers, making \manycall{} transparent to the programmer.
We intend to provide further tooling support in future work.
}

\subsection{Execution Model}

We have considered two alternative execution models, a return-oriented and a
call-oriented approach.
% Although both allow \ac{eBPF} programs to invoke system calls with
% programmable arguments and utilize the results,
\textcolor{plos21reviews}{\system uses a call-oriented approach because
  it offers better safety and efficiently implements our approachable
  programming model.}

% First, in an asynchronous, event-driven execution model the application
% registers \ac{eBPF} programs as handlers for system call results. The \ac{eBPF}
% programs can then request subsequent system calls either using a helper function
% or by returning a special value that describes the call to be executed.
% Second, in a synchronous execution model user space invokes a \ac{eBPF} program
% that runs in the context of the calling process. The program can then execute
% system call routines as subroutines and process their result when they return.

% APSys'21 Review #16D likes this: To help readers understand the idea easily,
% two or more alternative solutions are discussed in Section 3 and Section 4.
\paragraph{Return-Oriented}

In this execution model, \ac{eBPF} programs are executed in response to system
calls which are initiated asynchronously by the application.
This model has been proposed for integration in the Linux \texttt{io\_uring}
subsystem ~\cite{axboe:2020:plan,lwn:2021:ebpfiouring}.
With this model, the control flow is similar to \emph{return-oriented programming}~\cite{roemer:2012:acmtoiss}, an exploit technique where chunks are executed sequentially, and the \emph{return} instruction at the end of an instruction sequence leads to the execution of the following sequence.
For \systemCallX{} aggregation, such sequences are either \ac{eBPF}
programs, or system calls.
In this execution model, an \ac{eBPF} program can call a helper to request a subsequent system call which, on termination, triggers another \ac{eBPF} program.
It thus create a chain of system calls, interleaved with eBPF programs.
%This model has several disadvantages. First, termination of each \ac{eBPF} chunk is ensured, but the call chain may contain infinite loops. Second, the programmability of these chunks is not straight-forward as the control flow is encoded in return status codes. Third, the management of internal states for call chains is tricky as local variables have to be preserved along a call chain---all internal state needs concurrency control and garbage collection.
%
% The programs should allow for error handling, for example, aborting the system
% call burst by switching back into user space. The programs should also be able
% to decide freely which system call is to be executed next and which \ac{eBPF}
% program should receive that output of the subsequent call. If this can not be
% implemented it may also be sufficient to allow user space to define a single
% chain of syscalls (being the hottest code path) which are linked by predefined
% \ac{eBPF} programs. While executing the chain, any \ac{eBPF} program may request
% a switch back into user space if an exceptional condition is detected.

\begin{figure}[t]
  \lstdefinestyle{figmap}{
    language=C,
    breaklines=true,
    basicstyle=\ttfamily\footnotesize,
    keywordstyle=\ttfamily\footnotesize,
    commentstyle=\ttfamily\footnotesize\color{DarkGray},
    moredelim=**[is][\color{eBPF}\bfseries]{@}{@},
  }
  \begin{lstlisting}[style=figmap]
for (size_t i = 0; i < n @&& i < N@; i++) {
  fstat(fd[i], user_addr);    // Files, int fd[N];
  struct stat s;
  @copy_from_user(&s, sizeof(s), user_addr);@
  total += s.st_size;             // size_t total;
}
\end{lstlisting}
\caption{C Program for disk usage estimation, ported to \system by adding
  \texttt{\textbf{\textcolor{eBPF}{bold magenta}}} code. Error handling is
  omitted for brevity. $n < N$, where $N$ is a compile-time constant.}\label{fig:effort}
\end{figure}

\paragraph{Call-Oriented}

\begin{figure*}
  \begin{tikzpicture}[x=\xunit,y=\yunit]
    \tikzstyle{noflow} = [very thick]
    \tikzstyle{flow} = [noflow,decorate,decoration={snake,amplitude=1.3}, shorten >=0.0mm]
    \tikzstyle{flowVm} = [flow,color=eBPF]
    \tikzstyle{switch} = [-latex,thick,color=DarkGray]
    \tikzstyle{inactive} = [thick,color=Gray]
    \tikzstyle{labelEdge} = [thick,color=DarkGray]
    \tikzstyle{label} = [color=Black,font=\sffamily \footnotesize]

    \def\appMode{1.0}
    \def\appVm{2.0}
    \def\appKern{3.0}
    \def\modeVm{1.0}
    % TODO: 0.75
    \def\vmKern{1.0}
    \def\kernTime{0.5}

    \def\usRoot{0}
    \def\ebpfRoot{6}

    \def\modeSwitchDelay{0.5}
    \def\callDelay{0.1}

    \def\usZWidth{2}
    \def\usSyscallZWidth{1.2}
    \def\usYWidth{1.2}
    \def\usSyscallYWidth{1.2}
    \def\usXWidth{1.2}
    \def\usSyscallXWidth{1.2}
    \def\usWWidth{4}

    \def\cacheFactor{0.3}
    \def\vmAWidth{0.7}
    \def\syscallAWidth{\usSyscallZWidth}
    \def\vmBWidth{\usYWidth*\cacheFactor}
    \def\syscallBWidth{\usSyscallYWidth*\cacheFactor}
    \def\vmCWidth{\usXWidth*\cacheFactor}
    \def\syscallCWidth{\usSyscallXWidth*\cacheFactor}
    \def\vmDWidth{\vmAWidth}

    \node[anchor=east] (app) at (0,0) {Application};
    \node[anchor=east] (vm) at (0,-\appVm) {eBPF Program};
    \node[anchor=east] (kernel) at (0,-\appKern) {Kernel};

    % User space Syscall Z:
    \coordinate (usSyscallZApp) at ($(\usRoot,0)+(\usZWidth,0)$);
    \coordinate (usSyscallZKern) at ($(usSyscallZApp)+(\modeSwitchDelay,-\appKern)$);
    \coordinate (usSysretZKern) at ($(usSyscallZKern)+(\usSyscallZWidth,0)$);
    \coordinate (usSysretZApp) at ($(usSysretZKern)+(\modeSwitchDelay,\appKern)$);

    % User space Syscall Y:
    \coordinate (usSyscallYApp) at ($(usSysretZApp)+(\usYWidth,0)$);
    \coordinate (usSyscallYKern) at ($(usSyscallYApp)+(\modeSwitchDelay,-\appKern)$);
    \coordinate (usSysretYKern) at ($(usSyscallYKern)+(\usSyscallYWidth,0)$);
    \coordinate (usSysretYApp) at ($(usSysretYKern)+(\modeSwitchDelay,\appKern)$);

    % User space Syscall X:
    \coordinate (usSyscallXApp) at ($(usSysretYApp)+(\usXWidth,0)$);
    \coordinate (usSyscallXKern) at ($(usSyscallXApp)+(\modeSwitchDelay,-\appKern)$);
    \coordinate (usSysretXKern) at ($(usSyscallXKern)+(\usSyscallXWidth,0)$);
    \coordinate (usSysretXApp) at ($(usSysretXKern)+(\modeSwitchDelay,\appKern)$);

    \coordinate (ebpfRoot) at ($(usSysretXApp)+(\usWWidth,0)$);

    \coordinate (vmcallApp) at ($(ebpfRoot)$);
    \coordinate (vmcallVm) at ($(vmcallApp)+(+\modeSwitchDelay,-\appVm)$);
    % VM System Call A:
    \coordinate (syscallAVm) at ($(vmcallVm)+(\vmAWidth,0)$);
    \coordinate (syscallAKern) at ($(syscallAVm)+(\callDelay,-\vmKern)$);
    \coordinate (sysretAKern) at ($(syscallAKern)+(\syscallAWidth,0)$);
    \coordinate (sysretAVm) at ($(sysretAKern)+(\callDelay,\vmKern)$);
    % VM System Call B:
    \coordinate (syscallBVm) at ($(sysretAVm)+(\vmBWidth,0)$);
    \coordinate (syscallBKern) at ($(syscallBVm)+(\callDelay,-\vmKern)$);
    \coordinate (sysretBKern) at ($(syscallBKern)+(\syscallBWidth,0)$);
    \coordinate (sysretBVm) at ($(sysretBKern)+(\callDelay,\vmKern)$);
    % VM System Call C:
    \coordinate (syscallCVm) at ($(sysretBVm)+(\vmCWidth,0)$);
    \coordinate (syscallCKern) at ($(syscallCVm)+(\callDelay,-\vmKern)$);
    \coordinate (sysretCKern) at ($(syscallCKern)+(\syscallCWidth,0)$);
    \coordinate (sysretCVm) at ($(sysretCKern)+(\callDelay,\vmKern)$);
    % VM Return:
    \coordinate (vmretVm) at ($(sysretCVm)+(\vmDWidth,0)$);
    \coordinate (vmretApp) at ($(vmretVm)+(\modeSwitchDelay,\appVm)$);
    \coordinate (end) at ($(vmretApp)+(2,0)$);
    \coordinate (endVm) at ($(end)+(0,-\appVm)$);
    \coordinate (endKern) at ($(end)+(0,-\appKern)$);
    \coordinate (sepLabel) at ($(app.east)+(-0,-\appMode)$);

    \coordinate (tUsStart) at ($(usSyscallZApp)+(0,-\appKern)+(0,-\kernTime)$);
    \coordinate (tUsEnd) at ($(usSysretXApp)+(0,-\appKern)+(0,-\kernTime)$);
    \coordinate (tVmStart) at ($(vmcallApp)+(0,-\appKern)+(0,-\kernTime)$);
    \coordinate (tVmEnd) at ($(vmretApp)+(0,-\appKern)+(0,-\kernTime)$);

    % Incative Livelines (below other drawings):
    \draw[]
    (vm.east) edge[inactive] (vmcallVm)
    (vmcallApp) edge[inactive] (vmretApp)
    (kernel.east) edge[inactive] (usSyscallZKern)
    (sysretAKern) edge[inactive] (syscallBKern)
    (vmretVm) edge[inactive] (endVm)
    (sysretBKern) edge[inactive] (syscallCKern)
    (sysretCKern) edge[inactive] (endKern)
    ;

    % User/Kernel Mode Boundary:
    \node[anchor=south east,font=\sffamily \scriptsize] (unprivileged)
    at (sepLabel) {\textcolor{Black}{User Mode $\uparrow$}};
    \node[anchor=north east,font=\sffamily \scriptsize] (kernelMode)
    at (sepLabel) {\textcolor{Black}{Kernel Mode $\downarrow$}};
    \draw[] (kernelMode.north west) edge[labelEdge,thick,dashed] ($(end)+(0,-\appMode)$);

    % Time 1:
    \draw[]
    (tUsStart)
    edge[labelEdge,|->]
    node[label,below] {\textbf{$t$}}
    (tUsEnd)
    ;

    % Time 2:
    \draw[]
    (tVmStart) edge[labelEdge,|->] node[label,below] {\textbf{$t'$}} (tVmEnd);

    \draw[]
    (usSyscallZApp) edge[inactive] (usSysretZApp)
    (\usRoot,0) edge[flow,color=User] (usSyscallZApp)
    (usSyscallZApp) edge[switch] node[label,left,pos=0.16] {syscall} (usSyscallZKern)
    (usSyscallZKern) edge[flow,color=Kernel] (usSysretZKern)
    (usSysretZKern) edge[switch] (usSysretZApp)
    ;

    \draw[]
    (usSysretZKern) edge[inactive] (usSyscallYKern)
    (usSyscallYApp) edge[inactive] (usSysretYApp)
    (usSysretZApp) edge[flow,color=User] (usSyscallYApp)
    (usSyscallYApp) edge[switch] (usSyscallYKern)
    (usSyscallYKern) edge[flow,color=Kernel] (usSysretYKern)
    (usSysretYKern) edge[switch] (usSysretYApp)
    ;

    \draw[]
    (usSysretYKern) edge[inactive] (usSyscallXKern)
    (usSyscallXApp) edge[inactive] (usSysretXApp)
    (usSysretYApp) edge[flow,color=User] (usSyscallXApp)
    (usSyscallXApp) edge[switch] (usSyscallXKern)
    (usSyscallXKern) edge[flow,color=Kernel] (usSysretXKern)
    (usSysretXKern) edge[switch] (usSysretXApp)
    ;

    \draw[]
    (usSysretXKern) edge[inactive] (syscallAKern)
    (usSysretXApp) edge[flow,color=User] (vmcallApp)
    (vmcallApp) edge[switch] node[label,left,pos=0.25] {\system} (vmcallVm)
    (vmcallVm) edge[flowVm] (syscallAVm)
    % Syscall A:
    (syscallAVm) edge[switch] node[label,left] {kernel call} (syscallAKern)
    (syscallAVm) edge[inactive] (sysretAVm)
    (syscallAKern) edge[flow,color=Kernel] (sysretAKern)
    (sysretAKern) edge[switch] (sysretAVm)
    (sysretAVm) edge[flowVm] (syscallBVm)
    % Syscall B:
    (syscallBVm) edge[switch] (syscallBKern)
    (syscallBVm) edge[inactive] (sysretBVm)
    (syscallBKern) edge[flow,color=Kernel] (sysretBKern)
    (sysretBKern) edge[switch] (sysretBVm)
    (sysretBVm) edge[flowVm] (syscallCVm)
    % Syscall C:
    (syscallCVm) edge[switch] (syscallCKern)
    (syscallCVm) edge[inactive] (sysretCVm)
    (syscallCKern) edge[flow,color=Kernel] (sysretCKern)
    (sysretCKern) edge[switch] (sysretCVm)
    % VM Return:
    (sysretCVm) edge[flowVm] (vmretVm)
    (vmretVm) edge[switch] (vmretApp)
    (vmretApp) edge[flow,color=User] (end)
    ;

  \end{tikzpicture}
  \caption{Three identical synchronous system calls from user space (left) and
    using \system (right). The aggregated execution time $t'$ is smaller than $t$ as direct and indirect \transitionOverheads{} are reduced.}\label{fig:bpftasks}
\end{figure*}
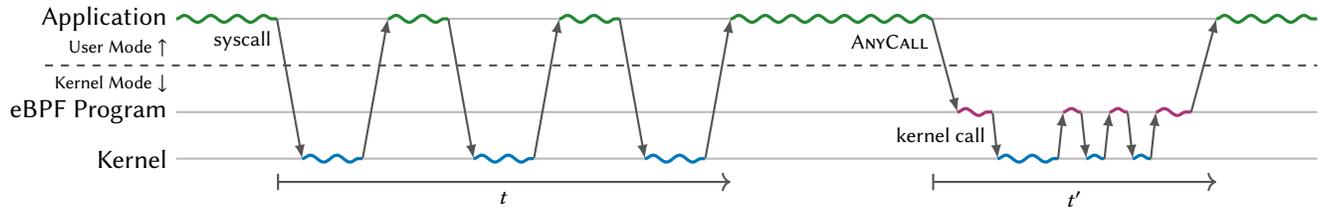

In a call-oriented execution model, \ac{eBPF} programs invoke
system calls synchronously, using stubs provided by the kernel and exposed to the eBPF bytecode. Equivalently to function calls,
the original \ac{eBPF} program receives the result upon \systemCallX{} completion and resumes execution. This model is displayed by example in \cref{fig:bpftasks}.

%\cref{fig:bpftasks} compares this to system calls from user space. Executing
%three consecutive systems calls from user space incurs a high direct and indirect
%cost. Executing the same three system calls using \ac{eBPF} incurs only one
%third of the direct mode switching overhead and speeds up both user and kernel
%code by not flushing processor state on each system call.

\paragraph{Discussion}

The return-oriented and call-oriented execution models differ in various
aspects, in particular, with respect to progress, safety, and programmability.

Regarding \emph{progress}, the return-oriented model verifies each individual \ac{eBPF} program for completion, but proving completion of the whole call chain is practically infeasible. This enables infinite loops involving alternating system calls and eBPF programs.
For \emph{safety}, the \systemCallX{} aggregation needs to manage its state, but protect it from other contexts. In the return-oriented model, the local state has to be preserved along a call chain, but protected from concurrent call chains. In consequence, this model demands for concurrency control. Furthermore, preserving the state correctly (including garbage collection at each end of the call chain) is non-trivial and cannot be safely ensured by the kernel.
Regarding \emph{programmability}, the call-oriented approach is straight-forward
as it resembles the way applications have interacted with the kernel
traditionally, which is consequently well-understood by many programmers. In
comparison, the return-oriented approach scatters control flow over multiple
chunks and embeds control flow decisions in helper calls.
In summary, \system implements the call-oriented execution model.
In consequence, \system guarantees completion of aggregated system calls, which limits flexibility, because the system-call aggregation as a whole is verified for completion. In consequence, only loops with provable bounds are allowed. As an example, file-system iteration is currently not possible, but future versions may support it via dedicated \ac{eBPF} helper functions.

\subsection{Access Privilege}

\textcolor{plos21reviews}{To use \system, an \ac{eBPF} program which aggregates the desired system calls
in an application-defined manner has to be loaded into the kernel.
As mentioned in \cref{sec:ebpf}, this is currently only available to privileged
or capability-granted processes. Trusted system services can therefore make
immediate use of our implementation using the \ac{eBPF} capability.}
% a privileged service can load them and pass them to subprograms by dropping privilges and having the fds still open
%To give untrusted applications access to trusted \ac{eBPF} bytecode (\eg
%generic, re-usable library \manycalls), a trusted
%\texttt{setuid}-based~\cite{mcilroy:1987:man} executable can load them and inherit
%them to the untrusted program using an open file descriptor.
% \item it is to be determined whether eBPF will be hardened to be useable by
% unprivileged processes, if there are usecases they may do it
However, several use-cases for \ac{eBPF} from unprivileged programs have been
discussed~\cite{lwn:2019:unprivileged,lwn:2021:seccompebpf}, besides \system.
Our approach may motivate further hardening efforts of the \ac{eBPF} subsystem
to allow safe execution of untrusted user-supplied code in the kernel address
space.

\section{Implementation}\label{sec:implementation}

This sections presents our \ac{eBPF}-based implementation of \manycall in Linux.
We discuss program loading, system call invocation, methods for
accessing user memory, and the programmer effort required for adoption of \system.

\begin{figure}
  \lstdefinestyle{figmap}{
    language=C,
    breaklines=true,
    basicstyle=\ttfamily\footnotesize,
    commentstyle=\ttfamily\footnotesize\color{DarkGray},
    moredelim=**[is][\color{Red}]{@}{@},
    moredelim=**[is][\color{Black}]{$}{$},
  }
  \begin{lstlisting}[style=figmap]
int *buf = map(user_addr, sizeof(int));
$if (!buf) return -1;$  // Check enforced by loader
*buf = 4;          // Valid, access within bounds
@*((long *) buf) = 4;@    // Invalid, out-of-bounds
$unmap(buf);$      // unmap-call enforced by loader
@*buf = 4;@     // Invalid, unmap() invalidated buf
\end{lstlisting}
\caption{\ac{eBPF} program accessing an integer at the virtual user address
  \texttt{user\_addr} using \texttt{map()} and \texttt{unmap()}. Invalid
  accesses to the memory area, which must be of constant size, are detected and
  prevented by the \ac{eBPF} static analyzer. Dereferencing \texttt{user\_addr}
  directly is not permitted.}\label{fig:map}
\end{figure}

\subsection{Loading and Invocation}

Our implementation adds a new \texttt{anycall} \ac{eBPF} program type to the
Linux kernel. Programs of the \texttt{anycall} type can be loaded using the
standard \texttt{bpf()} system call and therefore also integrate with existing
libraries that simplify \ac{eBPF} bytecode handling (\eg
\texttt{libbpf}~\cite{linux:2021:v511}). Usually, \ac{eBPF} programs are
attached to kernel events after loading and are then invoked asynchronously. To
enable our synchronous execution model, we create a new system call which invokes the \ac{eBPF} program referenced by a
file descriptor. Program registration is not required. The system call returns
once the \ac{eBPF} program has finished executing.

\subsection{Kernel Calls}

Our approach exposes system call implementations to \ac{eBPF} programs
as helper functions, called \emph{kernel calls}.
{\color{meeting210730}A similar but severely limited approach was recently introduced by the kernel developers in the context of code signing~\cite{lwn:2021:signed}.
%
% We use a dedicated system-call table as not every system call is available on
% every architecture. In consequence, only explicitly-{\color{red}permitted}
% system calls are supported by \system---which {\color{red} is no problem
%   because} \todo{why?} {\color{meeting210730} portable user code }.
%
In \ac{eBPF} bytecode, helpers are identified by numbers replaced with the
function address during compilation to machine code.} Therefore, the call-time overhead
is comparable to a subroutine call in native machine code.
Each kernel call is invoked using a dedicated helper function receiving up to
five arguments in \ac{eBPF} registers, which directly map to machine registers
on most architectures.
% Additional arguments are passed by reference using a memory
% array. \todo{relevant?}

\subsection{Access to User Memory}

Many system calls receive or return values by reference, and the existing
code-base of the Linux kernel expect that these references point to user space.
The numerous widely-scattered checks that system call arguments do not reference
kernel memory are required for security reasons, so disabling or bypassing them
is therefore not an option. Hence, \system requires a way to read and write user
memory in order to construct system call arguments and process their results.

{\color{plos21reviews}

  The \ac{eBPF} static analyzer already has the ability to track fixed-size
  dynamic memory allocations from helper functions.
% This is used to allow \ac{eBPF} to allocate and write {\color{meeting210730}
% records into ring buffers shared with user space}. This could be used like
% \texttt{malloc}/\texttt{free}~\cite{linux:2021:ringbuf},
  However, the available helpers allocate kernel memory which is unusable for
  system call arguments. Therefore, a new mechanism is required.}
%
% One solution we attempted used the \texttt{mmap} system call to create a new
% mapping in the address space of the calling process. This unused memory area
% could then be directly accessed by the \ac{eBPF} program similar to ringbuffer
% allocations. Unfortunately, accessing user space addresses while in kernel mode
% does not work on every architecture. In addition, one has to ensure the memory
% pages are present and not paged out which can not be guaranteed by passing flags
% to \texttt{mmap} (\eg \texttt{MAP\_POPULATE}).
%
We have considered two alternative solutions, one based on page-faults and the
other based on page pinning. \textcolor{plos21reviews}{\system implements the
  latter because it is more efficient when user memory is accessed repeatedly and requires fewer modifications to \system code.}
% Both give \ac{eBPF} the ability to access arbitrary
% user memory if it is safe, but they differ regarding their efficiency and API.

\paragraph{Page-Fault Handler}

% This solution is similar to the way most system call implementations access user
% space memory. First, \texttt{access\_ok()} checks if the memory area to be
% accessed is a vaild part of the process address space. Then, a page fault
% handler is activated using \texttt{user\_access\_begin()} which pages in any
% non-present pages in the area accessed by the kernel in the following (by
% default, accessing a user page which is not present in kernel mode would trigger
% a kernel panic). Finally, the memory is directly accessed in an
% architecture-specific way using \texttt{unsafe\_\{get|put\}\_user()}. On the
% AMD64 architecture, these functions simply dereference the passed pointers but
% special handling may be required on other architectures. When the access is
% complete, the page fault handler is deactivated by as call to
% \texttt{user\_access\_end()}.
This solution is similar to the way most system calls access data in user space.
Because user memory access is architecture-specific, helper functions are called
for every memory access. On x86, present user memory is accessed directly but
special operations are required on other architectures. {\color{meeting210730}If the access triggers a
page-fault, the helper functions return an error.}
%
% First, a page-fault handler for the memory area {\color{red}is activated} and
% {\color{red}it is checked if} the area is part of the process's address space.
% Second, {\color{red}special} helper functions are used to access parts of the
% area in an architecture-independent manner. On x86, present user memory can be
% accessed directly but special operations are required on other architectures.
% Finally, the page-fault handler {\color{red}is deactivated}.

% This interface could be made available to \ac{eBPF} as a helper function that
% receives an untyped user space address and its constant size and returns a
% context object after successfully running \texttt{access\_ok()} and
% \texttt{user\_access\_begin()} for the area. Thereafter \ac{eBPF} can invoke
% \texttt{unsafe\_\{get|put\}\_user()} on the area at certain offsets by invoking
% respective helper functions that receive the context object. Finally, the
% context object is invalidated and the page-fault handler removed by as helper
% call to \texttt{user\_access\_end()}. To avoid the overhead of frequent helper
% function calls the \ac{JIT} compiler could be modified to inline calls to
% \texttt{unsafe\_\{get|put\}\_user()} as it already happens for accesses to
% \ac{eBPF} maps.
This interface has the disadvantage that it requires an \ac{eBPF} helper call
for every access to user-space memory, we therefore did not pursue this approach.
Future work may extend the \ac{eBPF}-to-native compiler to inline these calls.
% to \ac{eBPF} {\color{red}map} accesses.
% \todo{map = ebpf-datenstruktur oder map()-funktion von unten?}

\paragraph{Page Pinning}

To avoid a helper call for every user memory access we have implemented an
alternative solution that only requires a \texttt{map()} and an \texttt{unmap()}
helper call for \textcolor{meeting210730}{each memory area}.
% Instead of relying on the page-fault
% handler {\color{red}for paging in the memory}, \texttt{map()} pins the pages
% into memory. Also, to allow direct access to the memory in a way portable across
% hardware architectures, the helper maps the physical memory into the
% {\color{red}virtual address range of the kernel}. This makes it directly
% accessible to \ac{eBPF} programs if the static analysis permits it.
We pin the pages into memory, thereby preventing page-faults, and map them to
kernel virtual addresses to guarantee direct, portable access to the memory from
\ac{eBPF}.
% The \texttt{map()} helper receives an untyped user pointer and its constant
% size, it returns a typed \ac{eBPF} pointer which can be directly dereferenced at
% offsets up to its constant size. This is safe because the helper function only
% returns a non-null pointer if the area is valid user memory. The static analysis
% ensures that (a) the returned \ac{eBPF} pointer is only dereferenced at valid
% offsets (the size is constant), (b) the \ac{eBPF} program performs an error
% check after calling \texttt{map()}, and (c) the memory mapping is removed before
% the program exits by calling \texttt{unmap()}. This deallocates the memory area
% in the view of the \ac{eBPF} static analyzer and therefore no subsequent
% accesses are permitted.
\textcolor{meeting210730}{The use of this interface and the checks performed by the
  \ac{eBPF} static analyzer are illustrated by example in \cref{fig:map}.}
%
% The interface enables direct, safe access to user memory areas of constant size,
% validated by the \ac{eBPF} static analyzer.

\section{Evaluation}
\label{sec:evaluation}

\newcommand{\manyCallVariant}[0]{\manycall variant}
\newcommand{\scvar}[0]{variant using traditional system calls}
\newcommand{\scvars}[0]{variants using traditional system calls}
\newcommand{\findManyCall}[0]{{\system}}
\newcommand{\sysBurst}[0]{{\texttt{sys-burst}}}
\newcommand{\getpid}[0]{\texttt{getpid()}}
\newcommand{\open}[0]{\texttt{open()}}
\newcommand{\close}[0]{\texttt{close()}}
\newcommand{\vectorOpen}[0]{vector \texttt{open()}}
\newcommand{\vectorClose}[0]{vector \texttt{close()}}

\begin{figure}[t]
  \includegraphics[width=8.5cm]{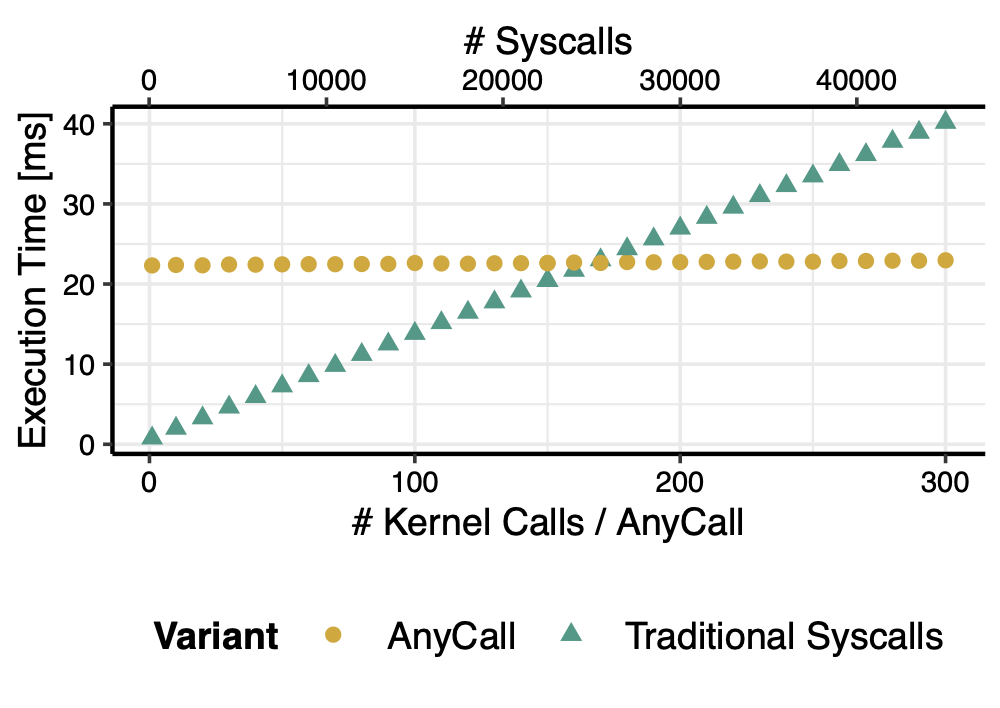}
  \caption{Execution time for an \manycall aggregating a variable number of
    \texttt{getpid()} system calls. The \manycall performs 1 to 300 kernel
    calls, and is invoked 150 times. The \scvar{} issues the equivalent number of
    system calls from user space.}\label{fig:getpidsize}
\end{figure}

We evaluate our implementation for the v5.11 Linux
kernel~\cite{linux:2021:v511} on an Intel Core i5-6260U
processor with two cores (four hardware threads) running at \SI{1.8}{\GHz}.
% faui4nuc$  cat /sys/devices/system/cpu/vulnerabilities/*
% Processor vulnerable
% Mitigation: PTE Inversion
% Mitigation: Clear CPU buffers; SMT vulnerable
% Mitigation: PTI
% Mitigation: Speculative Store Bypass disabled via prctl and seccomp
% Mitigation: usercopy/swapgs barriers and __user pointer sanitization
% Mitigation: Full generic retpoline, IBPB: conditional, IBRS_FW, STIBP: conditional, RSB filling
% Vulnerable: No microcode
% Not affected
%
%
The system configuration is left to the default values, therefore the vulnerability mitigations \ac{KPTI} and PTE Inversion are
active. % \Ac{DVFS} is disabled to not put the \manycall and \sysBurst{} variants from \cref{sec:realworld} at an advantage because they triggered higher CPU frequencies than the other, non-burst variants.
\textcolor{plos21reviews}{To make the evaluation reproducible, \ac{DVFS} is disabled.}

{\color{plos21reviews}
  We evaluate \system with \ac{KPTI} for multiple reasons. First, we expect that
  there is still a significant number of processors vulnerable to Meltdown
  deployed, simply because of the large number of affected devices when the
  vulnerability was disclosed. Second, as the discovery of new Meltdown-type
  attacks and development of respective mitigations is an ongoing
  process~\cite{musaev:2021:transient,hund:2013:kaslr,kemerlis:2012:kguard}, it
  is likely that the latency of \userKernelTransitions{} increases further. This
  is in line with the general development of an increasing latency of core
  \ac{OS} functionalities (\eg system calls)~\cite{ren:2019:sosp}.
  % This is supported by the fact that
  % \ac{KPTI} would have also been effective against multiple security
  % vulnerabilites discovered before
  % Meltdown~\cite{hund:2013:kaslr,kemerlis:2012:kguard}.
}
% ret2usr and KASLR bypass, also see
% https://github.com/a13xp0p0v/linux-kernel-defence-map/blob/master/linux-kernel-defence-map.svg

Each experiment is executed 20 times, but the first 10 iterations are considered warm-up iterations and thus discarded. In the following we report the average from the last 10 iterations. Error bars are omitted in the plots as the variation is not visible at the displayed scale.

\begin{figure}[t]
  \includegraphics[width=8.5cm]{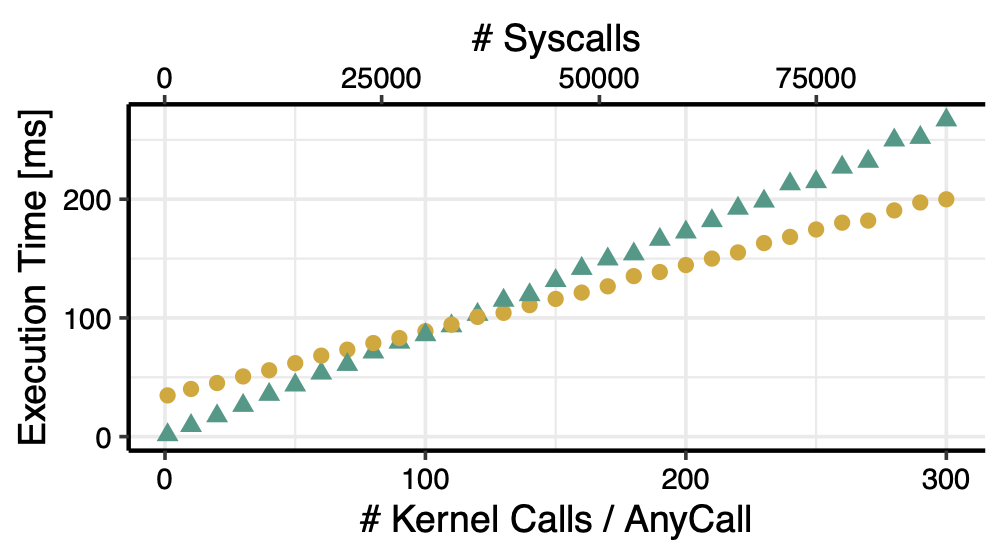}
  \caption{Execution time for two \manycalls performing a variable number of
    \open{} and \close{} kernel calls, and the equivalent variant using traditional system calls. The two \manycalls, each performing
    1 to 300 kernel calls, are both invoked 150 times.
    % The first \manycall
    % opens the set number of temporary files and the second \manycall closes each
    % file descriptor.
  }\label{fig:vector}
\end{figure}

\subsection{\textcolor{plos21reviews}{Microbenchmark}}

\textcolor{plos21reviews}{To analyze the degree to which \system reduces direct
  \userKernelTransition{} overheads,
  we have measured the execution time of \texttt{getpid()} performed with
  \system and in user space.
  % Because very little code executes in user- and
  % kernel space, the indirect transistion overhead is insignificant.
} The execution time of the \manyCallVariant{} includes the time for loading the
\ac{eBPF} program into the kernel.

% \paragraph{\manycall Size}

{\color{plos21reviews} Our motivating experiment in the introduction has
  demonstrated that it is not uncommon for real-world applications to perform
  \num{70000} to \num{726000} system calls per second. Therefore, our first
  experiment, displayed in \cref{fig:getpidsize}, executes between \num{150} and
  \num{45000} \texttt{getpid()} calls in user space or using \system (\manycalls
  with 1 to 300 kernel calls are invoked 150 times).} As expected, the time
required for static analysis and to-native compilation of the bytecode causes
the user-space variant to be faster if few system calls are performed inside the
program. However, \textcolor{plos21reviews}{if \system performs} the equivalent
of \num{25500} traditional system calls, it is faster than the user space
variant. Both variants show a linear relation between the executed work and the
execution time. The initial overhead \textcolor{plos21reviews}{of \system is}
\SI{22.34}{\ms} to prepare and load the \ac{eBPF} program into the kernel.
Thereafter, the \scvar{} requires \SI{131.8}{\us} for 150 \texttt{getpid()}
calls \textcolor{plos21reviews}{while \system uses} \SI{2.0}{\us}, it is
therefore 55 times faster.

Measuring the number of \ac{iTLB} misses reveals that each traditional system call
triggers one \ac{iTLB} load miss while kernel calls inside the \manycall{}
trigger none. The number of instructions per invocation is also reduced by
\SI{81}{\percent}. % from \num{34908} to \num{6493} per Manycall

% \paragraph{\manycall Uses}

% In the previous experiment we have varied the number of kernel calls performed
% by each \manycall and always invoked the loaded \ac{eBPF} program 150 times.
% The complementary experiment, that is, an \manycall that performs a constant number of kernel calls and is invoked between 1 and 300 times, shows almost identical results. Therefore, a plot is omitted.

%In
%another experiment we did the opposite: have a \manycall of constant size and
%invoke it between 1 and 300 times. We discovered that the resulting execution
%time is almost identical, therefore a plot displaying the results is omitted.

\subsection{Vector \manycalls}

\textcolor{plos21reviews}{In real-world applications, it is common to execute
  the same system call repeatedly on different data~\cite{debian:2020:manreadv}.
  Using \system, one can easily create such \emph{vector} versions for arbitrary
  system calls, even incorporating user-defined error handling.}
To demonstrate this, we have created vector versions of the \texttt{open()} and
\texttt{close()} system calls. Our \vectorOpen{} \manycall{} creates a requested
number of unnamed temporary files and stores the file descriptors into an array
(Opening named files is also possible by passing an array of strings to the
\manycall). The \vectorClose{} \manycall receives an array of file descriptors
and executes \close{} for each. In comparison to the \getpid{} experiment, more
memory is accessed in kernel and user space.

\Cref{fig:vector} displays the aggregated execution time for 150 invocations of
the vector \open{} and \close{} \manycalls (each invocation executes the
respective system call 1--300 times) and compares it to the equivalent \scvar.
The initial overhead to load the two \ac{eBPF} programs is \SI{33.65}{\ms}. The
execution time for the two \manycalls{} increases by \SI{0.56}{\ms} with each
processed file while the runtime of the variant doing traditional system calls
increases by \SI{0.87}{\ms} with each file. Therefore, the two \manycalls{} are
faster by \SI{36}{\percent}.
In comparison to our \getpid{} experiment execution-time difference is smaller,
as the \userKernelTransition{} overhead dominates the \getpid{} execution time.
However, the number of calls required to compensate for the loading overhead is
smaller because more time is saved per system call.
\textcolor{plos21reviews}{Approximately \num{16500} system calls suffice to
  justify the use of \system.}
% up to a certain size (where the initial time to get everything into the
%         cache does not matter anymore), we expect that the larger the
%         kernel/user code executed between/in system calls, the more quickly the
%         loading overhead is compensated
%
% itlb: 330 vs 0 misses per 300 system calls (openv + closev syscalls), at
%         least one misss per syscall saved
%
% indirect overhead is dominant because alsomost more memory is accessed
%         in user/kernel space
%
% eBPF uses more instructions initially, but uses fewer instructions in
%         each iteration (no register saving, dispatching, meltdown mitigations,
%         fewer timer interrupts because of the reduced runtime). The total number
%         of instructions is higher for eBPF in any of the displayed results,
%         however, eBPF might be at an advantage if a lot of system calls are
%         performed total. According to our model eBPF would use fewer
%         instructions at -216169+1584449x=143804624+1440457x which solves to
%         x=1000.2 -> when performing 150*2*1000=300000 syscalls the eBPF variant
%         will not only make better use of the caches but also use fewer
%         instructions

\subsection{Real-World File Searching Tool}\label{sec:realworld}

Many file types use \emph{magic values} at predefined offsets for identification.
To demonstrate that \system can speed up a real-world application, we have
applied it to a tool that filters files by such magic values.
For this tool, the list of files is received on standard input (generated by \texttt{find -type
  f}) and each file is opened, \texttt{seek}-ed, read and closed. If the contents read
from the offset match the magic value, the file path is written to standard
output.
In total, we have implemented four variants of the tool.
One uses \manycalls and three use traditional system calls:
\begin{description}
  \item[\system] To allow for a larger \ac{eBPF} program it
        is beneficial to read in a chunk of file paths and prepare an array of
        zero-terminated strings to be passed to the \ac{eBPF} program. The
        \ac{eBPF} program checks each path in the chunk using \texttt{open()},
        \texttt{lseek()}, \texttt{read()}, and \texttt{close()}. It
        conditionally calls \texttt{write()} to print matches.
  \item[{\texttt{sys-burst}}] executes the same algorithm as the
        \manyCallVariant{} but uses repeated traditional system calls.
  \item[{\texttt{sys}}] also checks the file contents using traditional system
        calls, but only reads in one file path at a time.
  \item[{\texttt{libc}}] checks the file contents using buffered \ac{IO}
	  based on C-library \texttt{FILE} pointers. Like \texttt{sys}, it processes one file path at a time.
\end{description}

\begin{figure}
  \includegraphics[width=8.5cm]{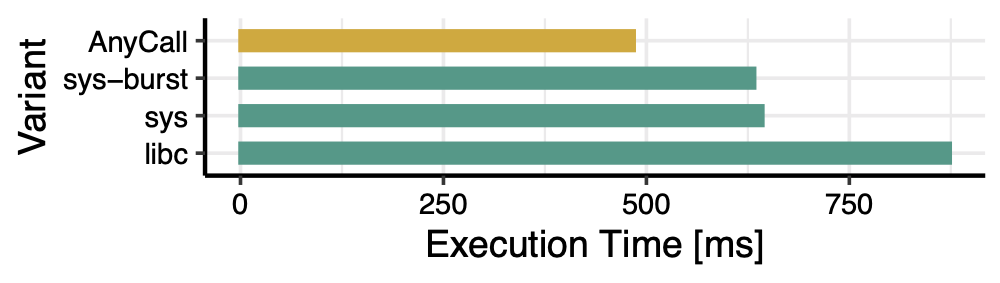}
  \caption{Execution time when searching all files with the \texttt{/bin/sh}
    shebang in the Linux source tree using \system and traditional system calls.}\label{fig:findmagic}
\end{figure}
The directory traversal by \texttt{find} and the processing of the input is
always performed in user space and included in the execution time.
\cref{fig:findmagic} displays the runtime when the Linux v5.0 sources are
checked for files with the \texttt{/bin/sh} shebang using the different
implementations.
Even though the \manycall variant issues only a single \manycall, it outperforms
the fastest user-space variant by \manyCallFindMagicSpeedup{}.

%   \item we have empirically determined the buffer sizes for the chunk-variants
The best chunk sizes for \findManyCall{} and \sysBurst{} were empirically
determined to be 512 and 1024.
%   \item user\_burst perfroms best at a buffer size of 1024*256byte (~= l2 cache)
%   \item bpf performs best at 512*256byte buffer size (larger was not possible
%         because the program became too complex)
Chunk sizes above 512 were not possible \textcolor{plos21reviews}{for \system,
  because} the \ac{eBPF} program became too large for static analysis.
\textcolor{plos21reviews}{However, \sysBurst{} is} already outperformed if the
chunk size is set to 4, which translates to only 24 kernel calls per
\manycall{}.

\section{Future Work}
\label{sec:futurework}

Future work will evaluate our \system implementation on systems without the
kernel mitigations for Meltdown and Spectre active
\cite{lipp:2018:meltdown,kocher:2019:spectre}.
%
%We will compare our call-oriented execution model to the asynchronous model of
%\texttt{io\_uring} maintainers as soon as it becomes
%available~\cite{axboe:2020:plan}.
%
We further intend to provide tool support for \manycalls.
Sections of a program to be run using an \manycall can be detected and transformed by a
compiler, making our solution transparent to the programmer.
We will demonstrate the advantages of \system on additional real-world
applications, in particular, database management systems, backup tools, and
other file-system utilities.
Generic, re-usable \manycalls, can be made available to untrusted applications
using \texttt{setuid} executables.

\section{Conclusion}
\label{sec:conclusion}

This paper has presented \system, which aggregates system calls and application-specific control logic, and executes as \ac{eBPF} bytecode in the Linux kernel.
\manycall maintains isolation while decoupling the number of \userKernelTransitions{} from the number of system calls.
Our run-time environment for \manycalls provides helper functions to access system calls from within \ac{eBPF}, as well as memory management.
Using a \getpid{} microbenchmark, we have demonstrated that kernel calls inside the
\manycall{} environment are up to 55 times faster than system calls from user space.
We have measured that vector \manycalls are \SI{36}{\percent} faster than the
user space equivalent,
and finally, have sped up a real-world file searching tool by
\manyCallFindMagicSpeedup{}.

\small
\begin{acks}
This work was partially funded by the Deutsche Forschungs\-gemeinschaft~(DFG) –- project number~465958100~(HO\,6277/1-1 and SCHR\,603/16-1) and by the Bundesministerium für Bildung und Forschung~(BMBF) -- project AI-NET-ANTILLAS (16KIS1315).
\end{acks}

\bibliographystyle{ACM-Reference-Format}
\bibliography{bib/references}

\end{document}